\documentclass[12pt,epsf]{article}
\pdfoutput=1
\usepackage{graphicx}
\usepackage{amsmath}
\usepackage{amsfonts}
\usepackage{cite}
\usepackage{color}
\begin{document}

\def\a{\alpha}
\def\b{\beta}
\def\c{\varepsilon}
\def\d{\delta}
\def\e{\epsilon}
\def\f{\phi}
\def\g{\gamma}
\def\h{\theta}
\def\k{\kappa}
\def\l{\lambda}
\def\m{\mu}
\def\n{\nu}
\def\p{\psi}
\def\q{\partial}
\def\r{\rho}
\def\s{\sigma}
\def\t{\tau}
\def\u{\upsilon}
\def\v{\varphi}
\def\w{\omega}
\def\x{\xi}
\def\y{\eta}
\def\z{\zeta}
\def\D{\Delta}
\def\G{\Gamma}
\def\H{\Theta}
\def\L{\Lambda}
\def\F{\Phi}
\def\P{\Psi}
\def\S{\Sigma}
\def\o{\over}
\def\beq{\begin{eqnarray}}
\def\eeq{\end{eqnarray}}
\newcommand{\gsim}{ \mathop{}_{\textstyle \sim}^{\textstyle >} }
\newcommand{\lsim}{ \mathop{}_{\textstyle \sim}^{\textstyle <} }
\newcommand{\vev}[1]{ \left\langle {#1} \right\rangle }
\newcommand{\bra}[1]{ \langle {#1} | }
\newcommand{\ket}[1]{ | {#1} \rangle }
\newcommand{\EV}{ {\rm eV} }
\newcommand{\KEV}{ {\rm keV} }
\newcommand{\MEV}{ {\rm MeV} }
\newcommand{\GEV}{ {\rm GeV} }
\newcommand{\TEV}{ {\rm TeV} }
\def\diag{\mathop{\rm diag}\nolimits}
\def\Spin{\mathop{\rm Spin}}
\def\SO{\mathop{\rm SO}}
\def\O{\mathop{\rm O}}
\def\SU{\mathop{\rm SU}}
\def\U{\mathop{\rm U}}
\def\Sp{\mathop{\rm Sp}}
\def\SL{\mathop{\rm SL}}
\def\tr{\mathop{\rm tr}}
\def\mpl{M_{PL}}

\def\IJMP{Int.~J.~Mod.~Phys. }
\def\MPL{Mod.~Phys.~Lett. }
\def\NP{Nucl.~Phys. }
\def\PL{Phys.~Lett. }
\def\PR{Phys.~Rev. }
\def\PRL{Phys.~Rev.~Lett. }
\def\PTP{Prog.~Theor.~Phys. }
\def\ZP{Z.~Phys. }


\baselineskip 0.7cm

\begin{titlepage}

\begin{flushright}
IPMU13-0189\\
ICRR-report-2013-9
\end{flushright}

\vskip 1.35cm
\begin{center}
{\bf A Closer Look at Gaugino Masses\\ in\\
Pure Gravity Mediation Model/Minimal Split SUSY Model
}

\vskip 1.2cm
Keisuke Harigaya$^1$, Masahiro Ibe$^{2,1}$ and Tsutomu T. Yanagida$^1$
\vskip 0.4cm
$^1${\it Kavli IPMU (WPI), TODIAS, University of Tokyo, Kashiwa 277-8583, Japan}\\
$^2${\it ICRR, University of Tokyo, Kashiwa 277-8582, Japan}
\vskip 1.5cm

\abstract{
We take a closer look at the gaugino masses in the context of pure
 gravity mediation models/minimal split SUSY models.
We see that the gaugino mass spectrum has a richer structure 
in the presence of vector-like matter fields even when they couple to 
the supersymmetry breaking sector only through Planck suppressed operators.
For example, the gluino mass can be much lighter than
in anomaly mediation, 
enhancing the detectability of the gluino at the LHC experiments.
The rich gaugino spectrum also allows new possibilities for dark matter scenarios
such as the bino-wino co-annihilation, bino-gluino co-annihilation, or even wino-gluino co-annihilation scenarios,
which affects future collider experiments as well as dark matter search experiments.
}
\end{center}
\end{titlepage}

\setcounter{page}{2}

\section{Introduction}
For decades, supersymmetry (SUSY) has been intensively studied  as a new fundamental ingredient 
of nature since it allows for a vast separation of low energy scales from high energy scales
such as the Planck scale or the scale of the Grand Unified Theory (GUT)\,\cite{MaianiLecture,Veltman:1980mj,Witten:1981nf,Kaul:1981wp}.
The unification of the gauge coupling constants at a very high energy scale also 
strongly motivates the supersymmetric standard model (SSM).
Weak-scale supersymmetry with the superparticle masses in the 
hundreds GeV to a TeV range, in particular, has been placed as top prospect
as it naturally solves the hierarchy problem while providing a good candidate for 
WIMP dark matter.

On the other hand, the supersymmetric standard models with heavier sfermions 
in the hundreds to thousands TeV range and the gauginos in the hundreds GeV
to a TeV range have also gained attention as attractive 
alternatives to the weak-scale supersymmetry\,\cite{Giudice:1998xp,Wells:2003tf}.
There, the neutral gauginos can be a good candidate of WIMP dark matter.
One of the biggest advantages of this class of models is that they do not 
require any singlet SUSY breaking fields (i.e. the Polonyi fields),
and hence, are free from the cosmological Polonyi problem\,\cite{Coughlan:1983ci,Ibe:2006am}.
These models are also free of the gravitino problem even for a very high reheating temperature\,\cite{Pagels:1981ke,Kawasaki:2004qu}, which is crucial for the successful thermal leptogenesis\,\cite{leptogenesis}.
The suppressions of flavor-changing neutral currents and $CP$ violating processes
are another important advantages of this class of models.

One apparent drawback of the heavy sfermion models is that the fine-tuning required 
for the weak-scale is one part of $10^{6}$--$10^{8}$.
In view of the above advantages, however, this drawback might be tolerable.
Alternatively, the anthropic principle\,\cite{Weinberg:1987dv}, 
which could be a manifestation of a large number 
of meta-stable vacua in string theory\,\cite{Bousso:2000xa,Kachru:2003aw,Susskind:2003kw,Denef:2004ze},
may explain the weak-scale to Planck scale hierarchy~\cite{Agrawal:1998xa,Jeltema:1999na}, 
which further supports the heavy sfermion models discussed 
in the context of split supersymmetry\,\cite{hep-th/0405159,hep-ph/0406088,hep-ph/0409232}.

We note in passing that the origin of the $\mu$-term has been a problem from the early stages of 
model building due to the absence of a singlet SUSY breaking 
fields\,\cite{Giudice:1998xp,Wells:2003tf}. 
This problem has been solved with a very simple model in Ref.\,\cite{Ibe:2006de} 
by utilizing a mechanism 
that generates a $\mu$-term similar in size to the gravitino mass from Planck suppressed interactions to the $R$-symmetry breaking sector\,\cite{Inoue:1991rk,Casas:1992mk}.%
\footnote{
The $\mu$-term which originates from $R$-symmetry breaking can be read off from the 
general formula of the $\mu$-term in Ref\,\cite{Giudice:1988yz}, and hence, it is sometimes regarded as 
a part of the (generalized) Giudice-Masiero mechanism.
The Giudice-Masiero mechanism uses, however, couplings between the Higgs sector and 
the SUSY-breaking Polonyi fields in the K\"ahler potential. 
Therefore, the mechanism for the $\mu$-term generation in the text 
is different from the Giudice-Masiero mechanism as stressed in\cite{Casas:1992mk}.
}
The model in Ref.\,\cite{Ibe:2006de} is now dubbed pure gravity mediation\,\cite{Ibe:2011aa,Ibe:2012hu}.
It should be noted that pure gravity mediation has an identical structure 
to minimal split SUSY
models\,\cite{minimalSplit,ArkaniHamed:2012gw}.%
\footnote{
In spread supersymmetry~\cite{Hall:2011jd}, the $\mu$-term generation by $R$
symmetry breaking is also mentioned as one of the possibilities.
In spread supersymmetry, however, a mediation scale $M_*$ other than the Planck scale is
introduced for the generation of the soft scalar masses, and hence the soft masses have a broader range than
in the case of pure gravity mediation models.
In Mini-split models~\cite{Arvanitaki:2012ps}, the Polonyi field is introduced in
order to generate the $\mu$ and the $B\mu$ term, and hence essentially
different from pure gravity mediation models/minimal split SUSY models.
}
In this paper, we use the name ``pure gravity mediation model''.

In pure gravity mediation models~\cite{Ibe:2011aa,Ibe:2012hu,minimalSplit,ArkaniHamed:2012gw}, 
the scalar bosons obtain SUSY breaking masses from the SUSY breaking sector via tree-level interactions 
of supergravity\,\cite{Nilles:1983ge}.
The $\mu$-term is,  as mentioned above,  generated from the $R$-symmetry breaking sector
via tree-level interactions of supergravity\,\cite{Inoue:1991rk,Casas:1992mk}.
The gaugino masses are, on the other hand, suppressed at the tree-level due to 
the absence of a singlet SUSY breaking field and are generated at the one-loop 
level mainly from the anomaly mediated SUSY breaking (AMSB)
contributions\,\cite{Giudice:1998xp,Dine:1992yw,Randall:1998uk}. 
The observed 126GeV Higgs boson mass~\cite{Aad:2012tfa,Chatrchyan:2012ufa} is explained with the gravitino mass
in the hundreds to thousands TeV range since ${\rm tan}\beta$, the ratio
of the two Higgs vacuum expectation values, 
is predicted to be ${\cal O}(1)$ in this class of model~\cite{Ibe:2011aa,Ibe:2012hu,minimalSplit,ArkaniHamed:2012gw}.

To date, the continuing absence of supersymmetric particles at the 
LHC is putting negative pressure on weak-scale supersymmetry 
and seems to favor heavier sfermion models.
Furthermore, the observed Higgs mass at around $126$\,GeV\,\cite{Aad:2012tfa,Chatrchyan:2012ufa} 
points to a sfermion mass scales above the tens of TeV range as anticipated in Ref.\,\cite{Okada:1990gg}.
In response to these results, the heavy sfermion models are gathering renewed 
attention from the view point of
collider and dark matter phenomenology~\cite{Alves:2011ug,Giudice:2011cg,Sato:2012xf,Bhattacherjee:2012ed,Hall:2012zp,Ibe:2012sx,Sato:2013bta},
low energy precision measurements~\cite{Hisano:2013cqa,Hisano:2013exa,Altmannshofer:2013lfa,Fuyuto:2013gla},
GUT or Planck scale theories~\cite{Linde:2011ja,Acharya:2012tw,Evans:2013lpa,Evans:2013dza}, and
cosmology~\cite{Nakayama:2012dw,Feldstein:2012bu,Harigaya:2012hn,Buchmuller:2013}. 

In this paper, we give a closer look at the gaugino masses in the
context of pure gravity mediation models.
The gaugino masses are the most important phenomenological parameters in the foreseeable future. 
In particular, we pay attention to the threshold corrections from vector-like matter fields.
As pointed out in Refs.~\cite{Nelson:2002sa,Hsieh:2006ig,Gupta:2012gu,Nakayama:2013uta},
these contributions can be comparable to the AMSB contributions even when
the vector-like matter fields couple to the SUSY breaking sector only
through Planck suppressed operators.
As we will see, the gluino mass can be much lighter than the AMSB predictions
for a given gravitino mass~\cite{ArkaniHamed:2012gw,Gupta:2012gu,Nakayama:2013uta},
enhancing the detectability of the
gluino at the LHC experiments.
We point out that the rich gaugino spectrum also allows
for new possibilities for dark matter
other than the standard thermal/non-thermal wino dark matter\,\cite{Moroi:1999zb,Gherghetta:1999sw,Hisano:2003ec,Ibe:2004tg},
including the bino-wino co-annihilation, the bino-gluino co-annihilation, or even the wino-gluino co-annihilation scenarios,
which affects future dark matter search experiments.
We emphasize the importance of the phases of the gaugino masses given by the
threshold corrections, which are not fully discussed in the literatures~\cite{ArkaniHamed:2012gw,Gupta:2012gu}.

\section{Gaugino Masses From Vector-Like Extra Matter}
\label{sec:vector}
In this section, we calculate the gaugino masses
when there exist vector-like matter fields $Q$ and $\bar{Q}$ which are charged 
under the SSM gauge symmetries.
We assume that $R$ charges of $Q\bar{Q}$ add up to $0$, 
so that the extra matter fields obtain 
a supersymmetric (Dirac) mass of order the gravitino mass from the $R$-breaking sector
as is the case for the higgs multiplets in pure gravity mediation models\,\cite{Inoue:1991rk,Casas:1992mk}.
We also assume that the extra matter fields couple to the SUSY breaking sector only
through Planck suppressed interactions similar to SSM matter fields 
in pure gravity mediation models. 

%

\subsection{SUSY breaking mass spectrum of extra matter}
To illustrate how the SUSY breaking effects show up in the mass spectrum of extra matter, 
let us consider the simplest SUSY breaking sector with the effective superpotential,
\begin{eqnarray}
W = \L^2 Z + m_{3/2} M_{PL}^2\ .
\end{eqnarray}
Here, $m_{3/2}$ denotes the gravitino mass representing the spontaneous (discrete) $R$-symmetry
breaking, and $M_{PL}$ the reduced Planck scale.
The SUSY breaking field $Z$ obtains an $F$-term vacuum expectation value (VEV)
of $F_Z = -\Lambda^2$, and the flat universe condition gives $\L^4 = 3 m_{3/2}^2M_{PL}^2$.%
\footnote{It is assumed that $|\vev{Z}|\ll \mpl$.}
In the followings, we take $m_{3/2}$ and $\L$ real and positive without loss of generality. 
As emphasized in the previous section, there is no singlet SUSY breaking field (i.e. the Polonyi field)
in the pure gravity mediation model. 
The SUSY breaking field $Z$ is assumed to be charged under some symmetries
at the Planck scale or to be a composite field generated at some dynamical scale much 
lower than the Planck scale.

Due to the vanishing $R$-charge of $Q\bar{Q}$, 
the extra matter couples to the above SUSY breaking sector via
the super- and K\"ahler potentials;
\begin{eqnarray}
 W &=& \Lambda^2 Z\left(1 +y
  \frac{Q\bar{Q}}{\mpl^2}\right) + m_{3/2}\mpl^2\left(1 + y'\frac{Q\bar{Q}}{\mpl^2}\right),\nonumber\\
 K &=& \lambda Q\bar{Q} + \lambda' Z^\dagger Z \frac{Q\bar{Q}}{\mpl^2}
  +~{\rm h.c.} + \cdots,
\end{eqnarray}
where
$y$, $y'$, $\lambda$ and $\lambda'$ are dimensionless coupling constants. 
It should be noted that we can eliminate one of $y$, $y'$ and $\lambda$ 
through the K\"ahler-Weyl transformation 
when we are only interested in the masses of the extra matter fields
(see also Ref.\,\cite{D'Eramo:2013mya}
for a related discussion).
In fact, by  using the K\"ahler-Weyl transformation,%
\footnote{
Since the K\"ahler-Weyl transformation involves chiral rotations of
fermion fields in chiral multiplets, it induces gauge kinetic
functions which are proportional to $\lambda Q\bar{Q}/\mpl^2$.
However, these terms do not contribute to gaugino masses. 
}
\begin{eqnarray}
 K\rightarrow K - \lambda Q\bar{Q} - \lambda^*Q^\dag \bar{Q}^\dag,~~~ W
  \rightarrow W {\rm exp}\left(\lambda Q\bar{Q}\right/\mpl^2),
\end{eqnarray}
the super- and K\"ahler potential can be rewritten as,
\begin{eqnarray}
 W' &=& \left(y'+ \lambda\right)m_{3/2} Q\bar{Q} + \sqrt{3}\left(y  +
       \lambda \right)m_{3/2} Z \frac{Q \bar{Q}}{\mpl} +
 \sqrt{3}m_{3/2}\mpl Z +
 \cdots,\nonumber\\
K' &=& \lambda' Z Z^\dag \frac{Q\bar{Q}}{\mpl^2} +~{\rm h.c.}\ .
\end{eqnarray}
Therefore, we obtain the supersymmetric Dirac mass, $M$, and the supersymmetry breaking 
mixing mass parameter, $b$,
\begin{eqnarray}
M &=& (y'+\lambda) m_{3/2}\ , \\
b &=& (3 y - y' + 2 \l - 3\l') m_{3/2}^2\ .
\end{eqnarray}
In deriving the expression of $b$, we have added up the contributions from the couplings to the SUSY breaking 
field and from the constant term in the superpotential through the supergravity interactions.%
\footnote{If $Q\bar{Q}$ couples to some flat directions, there also
exist contributions to the $b$ term by $F$ terms of the flat
directions~\cite{Pomarol:1999ie}. 
We assume, however, that $Q\bar{Q}$ do not couple to any flat directions.
}
As we expected, the parameters $y$, $y'$ and $\l$ show up in two combinations, 
$(y+\l)$ and $(y' + \l)$, although we keep all the parameters for later purpose.
As we will show, the phase of $b/M$ is a very important parameter for the gaugino masses.


\subsection{Gaugino masses from threshold corrections}
In order to calculate the threshold corrections, 
let us take the mass diagonalized basis for the extra matters. 
Here, it should be noted that in addition to the above mentioned SUSY breaking $b$-term, the scalar components of the 
extra matter generically obtain soft squared masses 
of order the gravitino mass just as the SSM matter fields do.
Thus, the mass terms of the scalar components $A,~{\bar A}$ and the
fermion components, $\psi,~\bar{\psi}$ are given by,
\begin{eqnarray}
 {\cal L}_{\rm mass-scalar} &=&
 - (|M|^2+\tilde{m}_A^2) |A|^2  - (|M^2| + \tilde{m}_{\bar{A}}^2) |\bar{A}|^2
  -\left( b A\bar{A} + {\rm h.c.}\right)
\nonumber\\
 &\equiv &- m_A^2 |A|^2  - m_{\bar{A}}^2 |\bar{A}|^2
  -\left( b A\bar{A} + {\rm h.c.}\right),\nonumber\\
 {\cal L}_{\rm mass-fermion} &=& - M \psi \bar{\psi} + {\rm h.c.}, 
\end{eqnarray}
respectively.
Here, $\tilde{m}_A^2$ and $\tilde{m}_{\bar{A}}^2$ denote the soft squared masses.
The mass terms of the scalar components are diagonalized by rotating the
fields,
\begin{eqnarray}
 \left(
\begin{array}{c}
 A_+\\
 A_-
\end{array}
\right)&=&
\left(
\begin{array}{cc}
 {\cos} \beta_Q & -e^{-i (\delta+\delta')}{\sin} \beta_Q \\
 e^{i (\delta + \delta') }{\sin} \beta_Q & {\cos} \beta_Q
\end{array}\right)
 \left(
\begin{array}{c}
 A\\
 \bar{A}^\dag
\end{array}
\right),\nonumber\\
\tan\beta_Q &=& \frac{m_{\bar{A}}^2-m_A^2 +
 \sqrt{(m_{\bar{A}}^2-m_A^2)^2 + 4 |b|^2}}{2 |b|}>0,\nonumber\\
\delta &=& {\rm
 arg}(b/M),~~
\delta' = {\rm
 arg}(M),
\end{eqnarray}
which leads to the mass eigenvalues, 
\begin{eqnarray}
 m_{\pm}^2 &=& \frac{1}{2}\left(m_A^2 + m_{\bar{A}}^2 \pm
			   \sqrt{\left(m_{\bar{A}}^2-m_A^2\right)^2 + 4
			   |b|^2} \right).
\end{eqnarray}

The one-loop threshold correction from the extra matter with the above 
mass spectrum yields the gaugino masses~\cite{Poppitz:1996xw},
\begin{eqnarray}
 \Delta m_{\lambda,{\rm threshold}}
= \frac{g^2}{16\pi^2} C_Q 2e^{i \delta} {\rm
  sin}2\beta_Q |M|
\left(
\frac{m_+^2}{m_+^2-|M^2|}
{\rm ln}\frac{m_+^2}{|M|^2}
-
\frac{m_-^2}{|M|^2-m_-^2}
{\rm ln}\frac{|M|^2}{m_-^2}
\right), 
\end{eqnarray}
at the renormalization scale just below their threshold. 
Here, $C_Q$ is a Dynkin index of $Q$, which is normalized to be $1/2$
for a fundamental representation, and $g$ the gauge coupling constant 
evaluated at around the scale of the extra matter.
By adding the AMSB effects of the extra matter, 
$\Delta m_{\lambda,{\rm AMSB}} = g^2/(16\pi^2) 2 C_Q m_{3/2}$,
we obtain the final result,%
\footnote{This formula can be applied to any cases, no matter the
origin of the Dirac mass, $b$ term, and soft squared mass terms.}
\begin{eqnarray}
\label{eq:deltam vector}
 \Delta m_{\lambda} 
&=& \frac{g^2}{16\pi^2} 2C_Q \left(e^{i \delta} {\rm
  sin}2\beta_Q |M|
\left(
\frac{m_+^2}{m_+^2-|M^2|}
{\rm ln}\frac{m_+^2}{|M|^2}
-
\frac{m_-^2}{|M|^2-m_-^2}
{\rm ln}\frac{|M|^2}{m_-^2}
\right)
+m_{3/2}
\right).\nonumber\\
\end{eqnarray}

Before closing this subsection, several comments are in order.
First, it can be proven that
\begin{eqnarray}
\frac{m_+^2}{m_+^2-|M^2|}
{\rm ln}\frac{m_+^2}{|M|^2}
-
\frac{m_-^2}{|M|^2-m_-^2}
{\rm ln}\frac{|M|^2}{m_-^2}
>0.
\end{eqnarray}
Therefore, the phase of the gaugino mass contributed from the threshold
correction is always determined by the phase of $b/M$.

Secondly, let us take the limit of small soft squared masses, i.e. $\tilde{m}_A^2,~\tilde{m}_{\bar{A}}^2\ll |M|^2$.
In this limit, the diagonalized scalar masses and mixing angle are reduced to
\begin{eqnarray}
 m_{\pm}^2 = |M|^2\pm |b|,~~
 {\rm tan}\beta_Q = 1.
\end{eqnarray}
With this mass spectrum, Eq.~(\ref{eq:deltam vector}) is also reduced to
\begin{eqnarray}
\label{eq:deltam nosquared}
 \Delta m_\lambda &=& \frac{g^2}{16\pi^2}2 C_Q\left[
\frac{b}{M} F(|b/M^2|)+m_{3/2}
\right],\nonumber\\
F(x)&\equiv& \frac{1+x}{x^2}{\rm ln}(1+x) +\frac{1-x}{x^2}{\rm ln}(1-x).
\end{eqnarray}
In order for the scalar components of $Q\bar{Q}$ not to be tachyonic, 
the $b$ term should satisfy $|b|<|M|^2$, where 
the function $F$ takes values between $1$ to ${\ln}(4)\simeq1.4$.

Thirdly, let us consider the limit of $|y'|\gg 1$. 
In this case, the spectrum for the extra matter is similar to the case with a large Dirac mass term in the super-potential.
Therefore, we expect that $Q\bar{Q}$ decouples and $\Delta m_{\lambda} =0$ as expected from
the ultraviolet insensitivity properties of the AMSB spectrum.
Actually, since the Dirac mass term and the $b$ term are given by $M=y'm_{3/2}$ and $b/M = - m_{3/2}$, 
and the soft squared mass terms are negligible, 
we obtain $\Delta m_\lambda=0$ from Eq.~(\ref{eq:deltam nosquared}).

Finally, let us take the limit of $|\lambda| \gg 1$, where the Dirac mass term and
the $b$ term are given by $M = \lambda m_{3/2}$ and $b/M = 2m_{3/2}$.
The soft squared mass terms are negligible and we obtain
\begin{eqnarray}
 \Delta m_\lambda = \frac{g^2}{16\pi^2} 6 C_Q m_{3/2}.
\end{eqnarray}
The AMSB effect and the threshold correction contribute to gaugino
masses additively.

\subsection{The SSM gaugino masses}
The AMSB contributions from the minimal supersymmetric standard model (MSSM) fields to the gluino mass $M_3$, the wino
mass $M_2$, and the bino mass $M_1$ are given by
\begin{eqnarray}
\label{eq:MSSM}
 M_{3}^{\rm AMSB}=-\frac{g_3^2}{16\pi^2}3m_{3/2},~~M_2^{\rm
  AMSB}=\frac{g_2^2}{16\pi^2}m_{3/2},~~M_1^{\rm AMSB}=\frac{g_1^2}{16\pi^2}\frac{33}{5}m_{3/2},
\end{eqnarray}
at around the renormalization scale $m_{3/2}$.
Besides the contribution, the wino and the bino obtain threshold corrections from 
the higgsino as in Ref.~\cite{Giudice:1998xp},
\begin{eqnarray}
\label{eq:Higgsino}
 \Delta M_2 = \frac{g_2^2}{16\pi^2}L,~~\Delta M_1 =
  \frac{g_1^2}{16\pi^2}\frac{3}{5}L,\nonumber\\
L\equiv \mu ~{\rm sin }2\beta \frac{m_A^2}{|\mu|^2-m_A^2}{\rm ln}\frac{|\mu|^2}{m_A^2},
\end{eqnarray}
where $\beta$ is defined by the ratio of the vacuum expectation value of the
up type higgs to that of the down type higgs, ${\rm tan}\beta=v_u/v_d$, and
$m_A^2$ is the mass squared of heavy higgs bosons.
As emphasized in Ref.\,\cite{Ibe:2012hu}, the higgsino threshold corrections
can be comparable to AMSB contributions in large sections of parameter space in pure gravity mediation models 
since it predicts $\mu \sim m_A = {\cal O}(m_{3/2})$ and $\tan\b = {\cal O}(1)$.

As we have found, the extra-matter fields give additional contributions to the gaugino masses
as in Eq.\,(\ref{eq:deltam vector}).
The physical gaugino masses are evaluated by solving renormalization equations,
\begin{eqnarray}
\label{eq:renormalization eq}
 \frac{{\rm dln}M_i(\mu)}{{\rm dln}\mu} &=&
  -\frac{g_i^2(\mu)}{8\pi^2}b_i,~~~(b_1,b_2,b_3)= (0,6,9),\nonumber\\
M_i(M_{i,{\rm phys}}) &=& M_{i,{\rm phys}}. 
\end{eqnarray}
Boundary conditions are given by the sum of the contributions given in
Eqs.\,(\ref{eq:MSSM}), (\ref{eq:Higgsino}) and \,(\ref{eq:deltam
vector}) evaluated at around the renormalization scale $m_{3/2}$.


In the following, let us suppose that the vector-like matter fields belong to $SU(5)$ GUT
multiplets so that coupling unification is preserved.
In this case, the contributions from the vector-like matter fields are given by,
\begin{eqnarray}
\label{eq:gaugino mass parametrization}
 \Delta M_i = \frac{g_i^2}{16\pi^2} e^{i\gamma} N_{\rm eff}m_{3/2},
\end{eqnarray}
and hence, satisfy the so-called GUT relation.
The definition of $N_{\rm eff}$ can be understood by comparing Eqs.\,(\ref{eq:deltam vector})
and (\ref{eq:gaugino mass parametrization}).%
\footnote{$N_{\rm eff}$ is not identical to the number of flavors,
$\sum_Q 2C_Q$.}
It should be noted that $N_{\rm eff}$
can be rather large
either from small $m_-^2$ or from many extra matter fields.
As we have discussed, the phase of $b/M$ is a free parameter, 
and hence, we take $\gamma$ as a free parameter.
This is a crucial difference from axion models, where
there is no phase freedom as reviewed in the appendix.\,\ref{sec:axion}.

Let us comment on $CP$ violations from the phase of the gaugino masses.
First, we assume that some flavor symmetry controls the soft squared
mass terms so that they are nearly diagonal, since otherwise constraints
from the $K^0-\bar{K}^0$ mixing suggest that the soft squared mass
terms are larger than ${\cal O}(1000)$
TeV~\cite{Gabbiani:1996hi,Bhattacherjee:2012ed}, even if $\gamma=0$.
Under this assumption, a one-loop contribution to
the neutron electric dipole moment is much smaller than
the experimental upper bound~\cite{Fuyuto:2013gla}.
A two loop Barr-Zee type contribution, which dominates over
the one loop contribution for large soft squared mass terms,
is also far smaller than the experimental upper bound for $\mu={\cal
O}(100)$ TeV~\cite{ArkaniHamed:2004yi}.

\begin{figure}[tb]
\begin{tabular}{cc}
\begin{minipage}{0.5\hsize}
\begin{center}
  \includegraphics[width=0.8\linewidth]{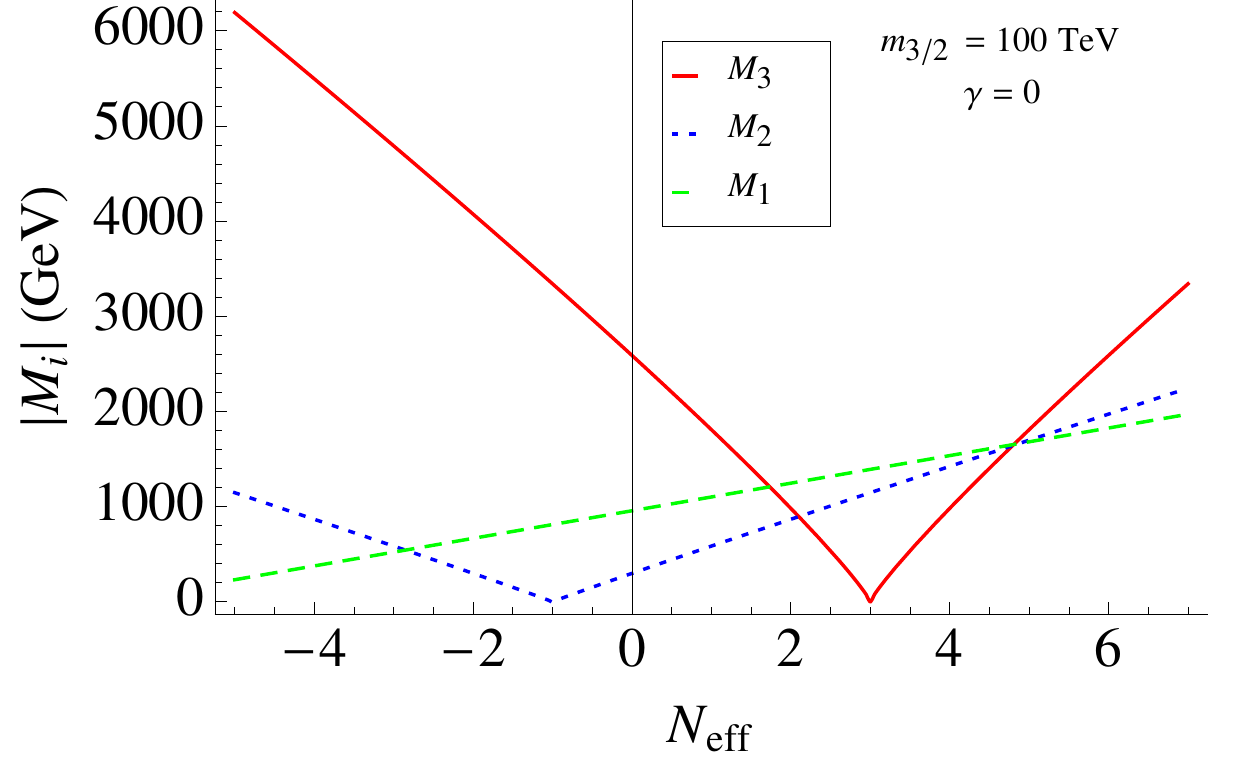}
\end{center}
\end{minipage}
 &
\begin{minipage}{0.5\hsize}
\begin{center}
  \includegraphics[width=0.8\linewidth]{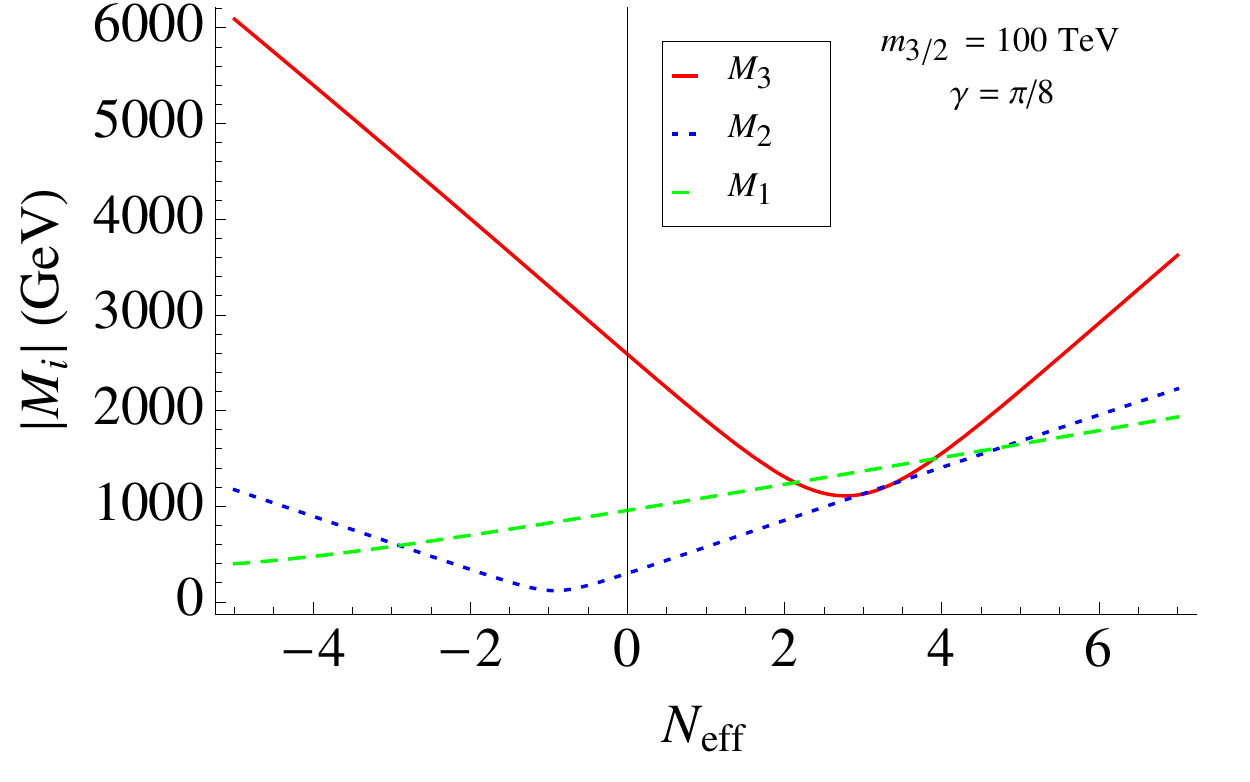}
\end{center}
\end{minipage}
\\
\begin{minipage}{0.5\hsize}
\begin{center}
  \includegraphics[width=0.8\linewidth]{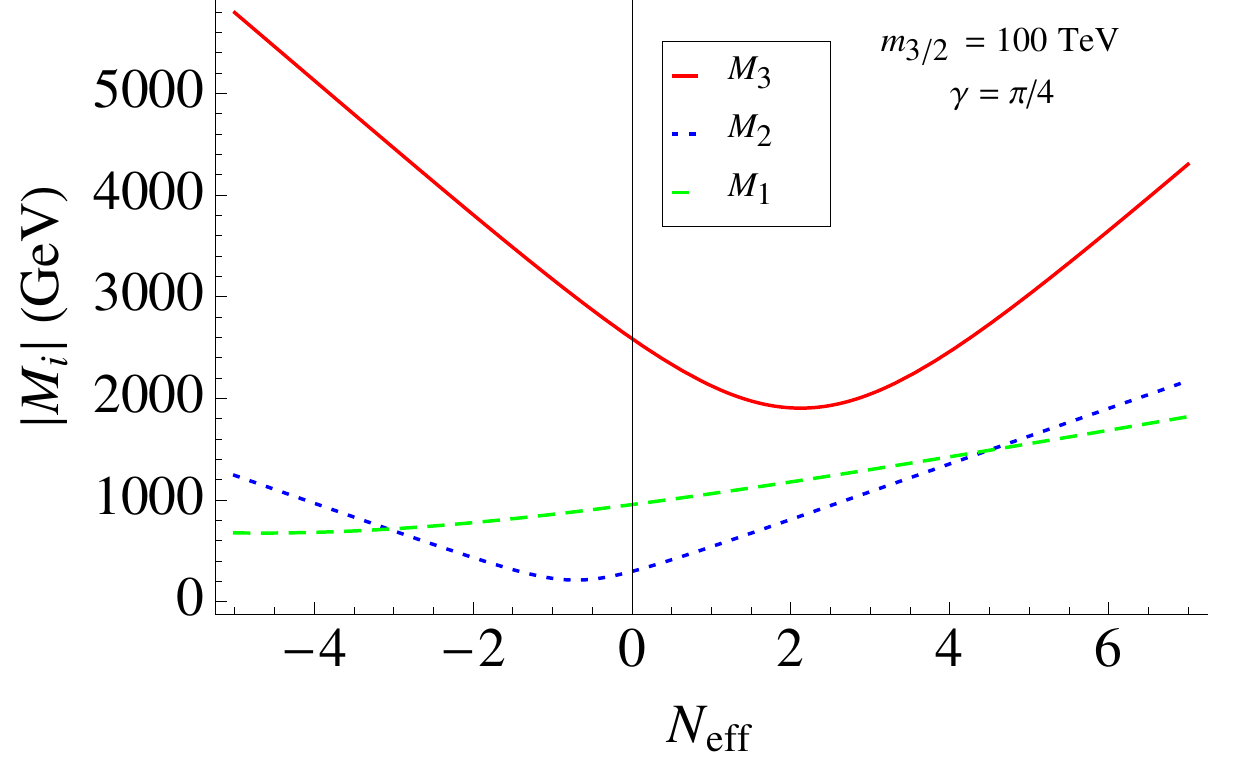}
\end{center}
\end{minipage}
 &
\begin{minipage}{0.5\hsize}
\begin{center}
  \includegraphics[width=0.8\linewidth]{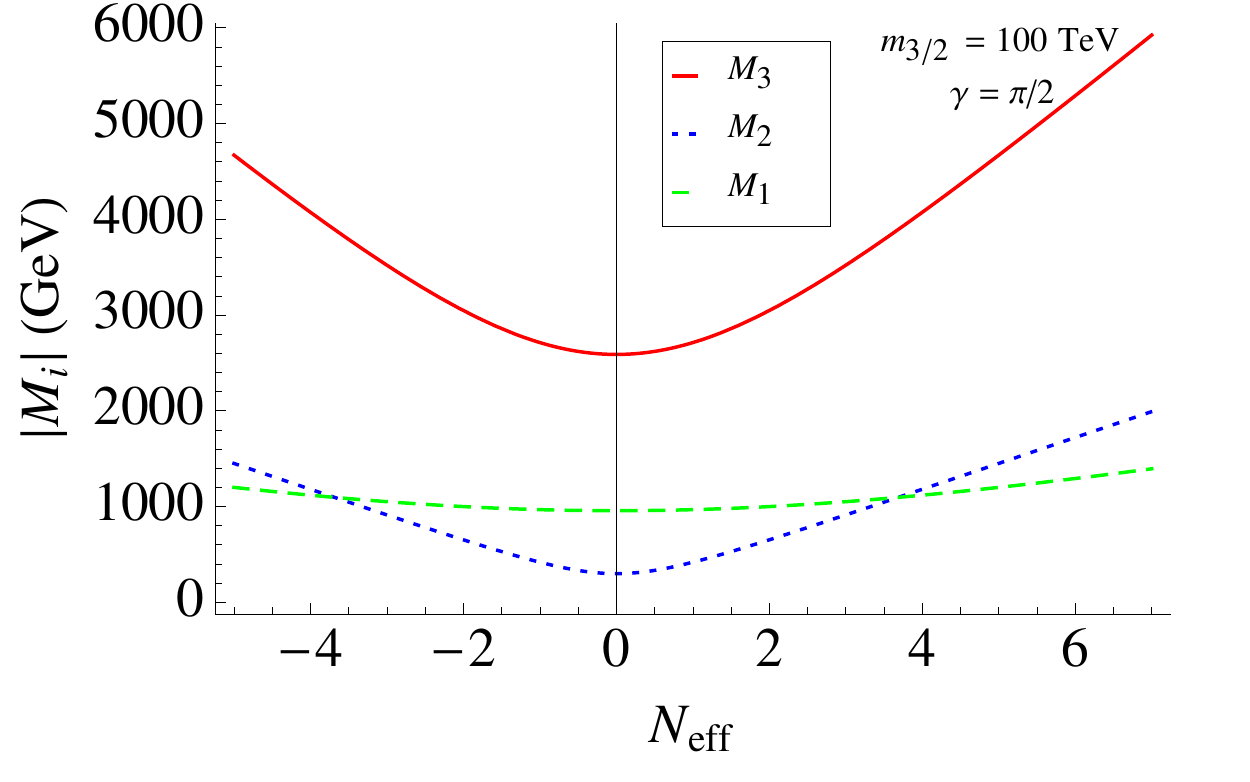}
\end{center}
\end{minipage}
\end{tabular}
\caption{\sl \small
The gluino, wino, and bino masses for $m_{3/2} = 100$ TeV with the threshold
 corrections from the extra vector-like matter in Eq.\,(\ref{eq:gaugino mass parametrization}).
We have neglected the higgsino threshold correction, for simplicity, i.e. $L=0$.
}
\label{fig:vector-mass}
\end{figure}

In Figure~\ref{fig:vector-mass}, we show the physical gaugino masses 
in the presence of the extra matter fields for $m_{3/2}=100$ TeV
as a function of $N_{\rm eff}$ for given values of $\gamma$.
Here, we have neglected the higgsino threshold correction for
simplicity, i.e. $L=0$.
It can be seen that the gluino mass can be much lighter than that
predicted in pure AMSB.
The figures also show that the gaugino mass spectrum strongly depends on the phase $\gamma$.
It should be noted that it is even possible for all three gauginos to be
degenerate for $\gamma \simeq 0$ and $N_{\rm eff} \simeq 4-5$.
This is caused by the fact that the MSSM contributions to the gluino mass is negative
while those to the wino and the bino masses are positive.
Thus, the addition of the extra matter contributions satisfying the GUT relation
can reduce the gluino mass while increasing the wino and bino masses.

\begin{figure}[tb]
\begin{tabular}{cc}
\begin{minipage}{0.5\hsize}
\begin{center}
  \includegraphics[width=0.8\linewidth]{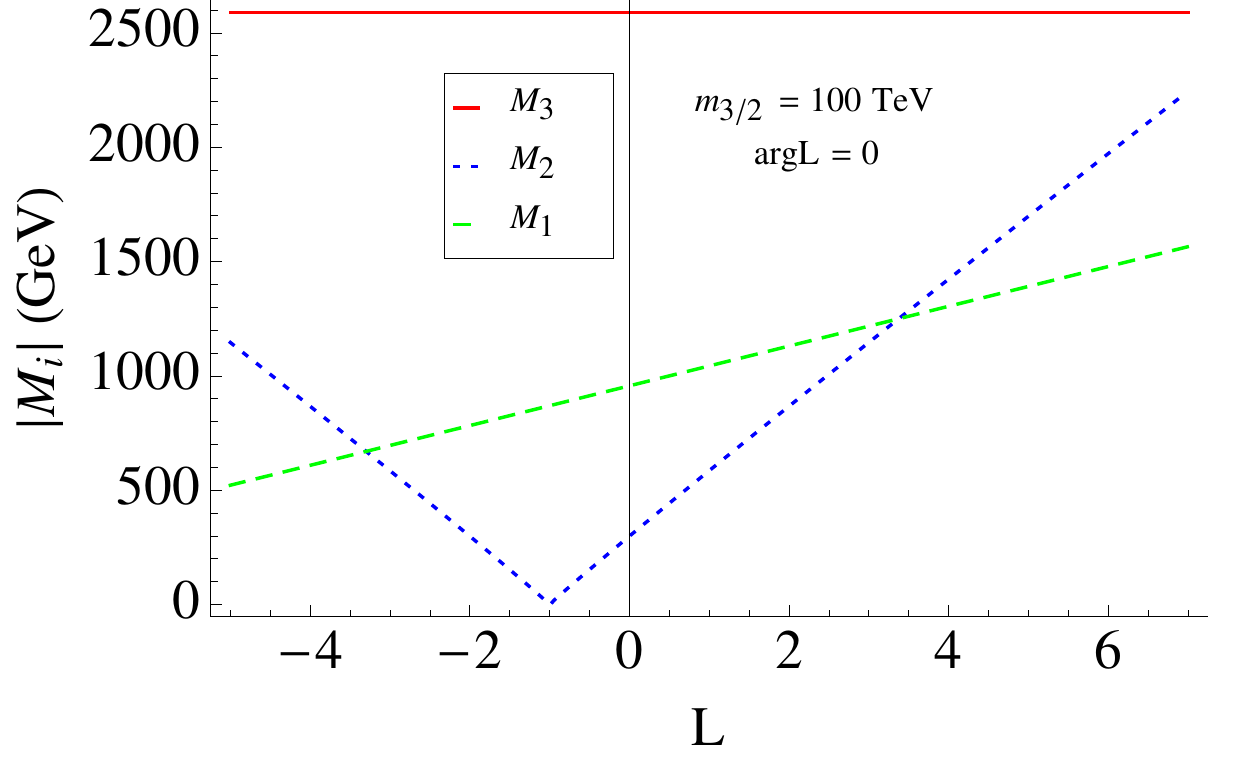}
\end{center}
\end{minipage}
 &
\begin{minipage}{0.5\hsize}
\begin{center}
  \includegraphics[width=0.8\linewidth]{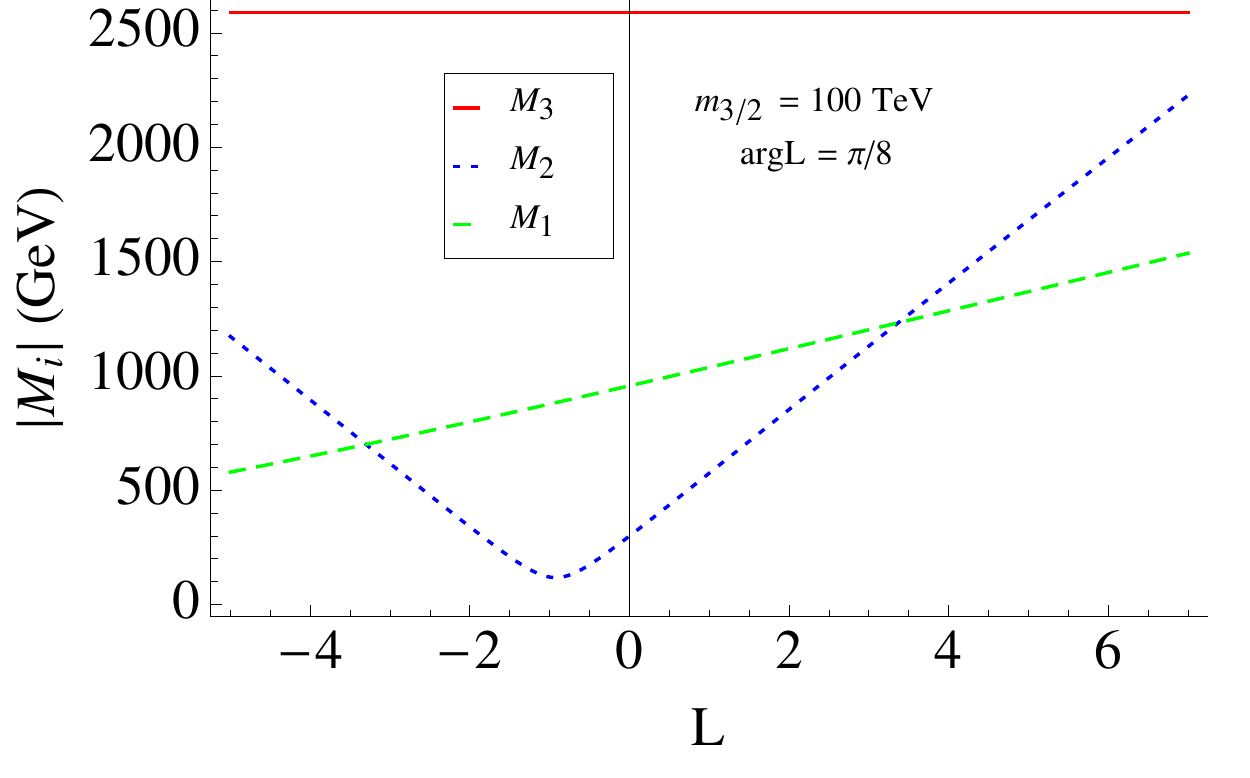}
\end{center}
\end{minipage}
\\
\begin{minipage}{0.5\hsize}
\begin{center}
  \includegraphics[width=0.8\linewidth]{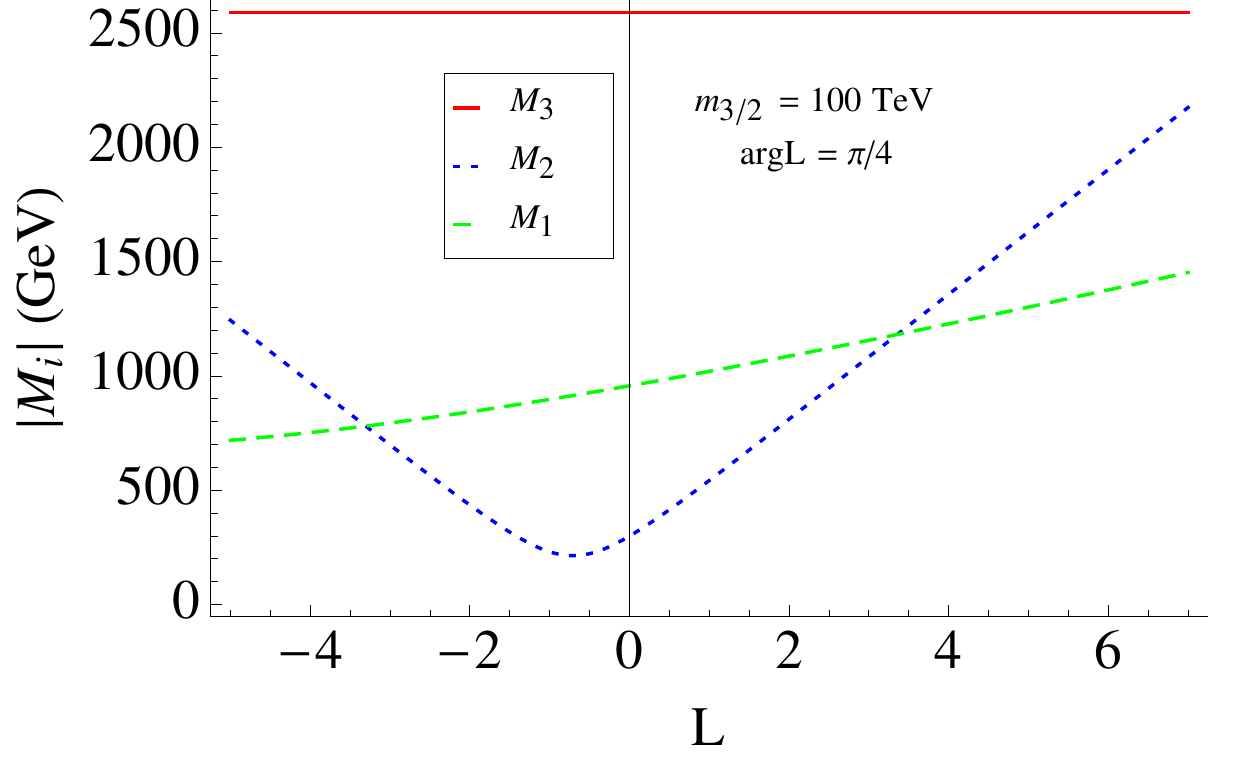}
\end{center}
\end{minipage}
 &
\begin{minipage}{0.5\hsize}
\begin{center}
  \includegraphics[width=0.8\linewidth]{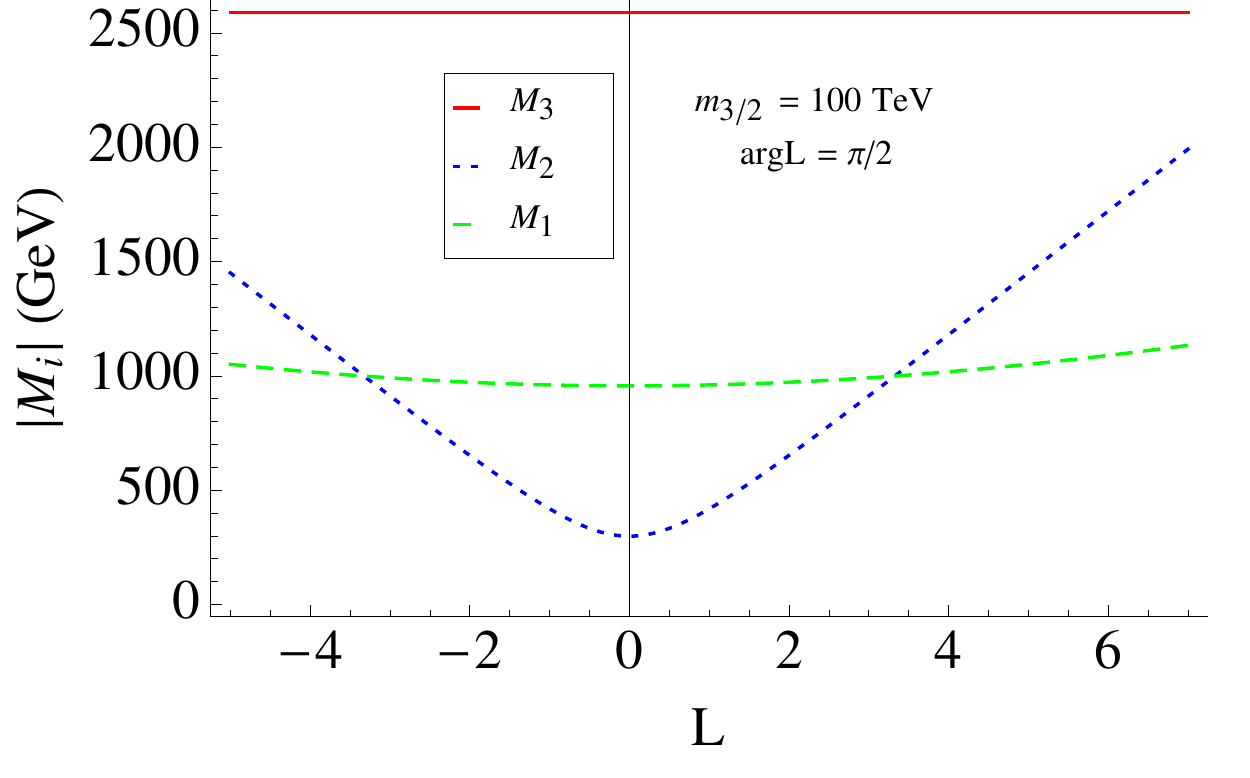}
\end{center}
\end{minipage}
\end{tabular}
\caption{\sl \small
The gluino, wino, and bino masses for $m_{3/2} = 100$ TeV with the higgsino
 threshold corrections.
}
\label{fig:pure-mass}
\end{figure}

For a comparison, in Figure~\ref{fig:pure-mass}, we show the physical gaugino masses 
when there are no extra vector-like matter fields, i.e. $N_{\rm eff} =0 $, 
but the higgsino threshold corrections are included, i.e. $L\neq 0$.
Since the gluino mass does not receive any corrections from the higgsino
threshold, it is difficult for the gauginos to have a rather degenerated spectrum.

In Figures~\ref{fig:50TeV}, \ref{fig:100TeV} and \ref{fig:300TeV}, we also show some
parameter regions which are phenomenologically distinctive including:
\begin{list}%
 {$\bullet$} 
 {} 
 \item 
Regions in which the bino/gluino is the lightest supersymmetric particle (LSP)
 \item
Regions in which the thermal abundance of the LSP is larger than
       the measured abundance of the cold dark matter, $\Omega_ch^2 =
      0.1199\pm 0.0027$~\cite{Ade:2013zuv}
 \item
Regions which are excluded by charged wino searches~\cite{charged-wino}
 \item
Regions which are excluded by the gluino searches~\cite{gluino}
 \item
Contour plots of the mass of the wino, $M_2$
 \item
Contour plots of the ratio $M_3/M_{\rm LSP}$ 
\end{list}
for various $m_{3/2}$,  $L$ and $N_{\rm eff}$.
For the calculation of the thermal abundance of the LSP, we have utilized
micrOMEGAs 3.2~\cite{Belanger:2013oya}. 
For simplicity, we have neglected the Sommerfeld enhancement
of the annihilation cross sections of the wino and gluino~\cite{Hisano:2006nn}.%
\footnote{
If one includes the effect of the Sommerfeld enhancement, the widths of the bino LSP
bands discussed later are modified due to the change in the wino or
gluino annihilation cross section.
For a qualitative discussion, see Ref.~\cite{HKM}.
Furthermore, the regions where the thermal abundance of the wino is consistent with the observed
dark matter abundance are shifted to higher mass scale regions,
$M_2\simeq 3$ TeV~\cite{Hisano:2006nn}.
}

Let us first examine Figure~\ref{fig:50TeV}, where $m_{3/2}=50$ TeV.
Usually, when the bino is the LSP, the thermal abundance of the LSP
exceeds the observed value (the left most and the right most regions). 
However, there exist bands
in which the bino is the LSP but the thermal abundance of
the LSP does not exceed the observed value due to the co-annihilations
with the wino~\cite{BirkedalHansen:2001is} or gluino~\cite{Profumo:2004wk}.
In these regions, it is possible that the bino is the dark matter and
is difficult to be observed through direct/indirect detections, because
the bino interactions are suppressed.

As the figures show, the ratio between the gluino mass and the LSP mass,
 $M_3/M_{\rm LSP}$, strongly depends on the phase $\gamma$. 
If $M_3/M_{\rm LSP}$ is close to one, for example,
the decay products of gluinos are soft. 
Thus, in such cases, even if gluinos are copiously produced at the LHC, 
the search for SUSY events requires initial state radiation~\cite{Alwall:2009zu,Bhattacherjee:2013wna}.
Therefore, the phase parameter $\gamma$  is not only important for the dark matter properties
but also important for collider searches.

In much of the right portion of Figure~\ref{fig:300TeV}, the wino is the LSP and too
heavy and the thermal abundance of the wino is larger than the observed
value. 
However, due to co-annihilation with the gluino, there exist a region in
which the thermal abundance is still smaller than the observed value
(the lower right panel, a white band between the colored regions).
This extends the thermal wino dark matter regions allowing heavier winos.

Let us comment about constraints on the dark matter from indirect dark matter searches.
As discussed in Refs.\cite{Cohen:2013ama,Fan:2013faa}, the wino LSP is constrained by indirect dark matter searches, 
in particular, by the gamma-ray searches in the Fermi and H.E.S.S. experiments from the Galactic center 
and the dwarf Spheroidal galaxies.
The constraints from the gamma-ray searches from the Galactic center,
however, suffer from large ambiguities 
of a dark matter profile (see e.g. Ref.\,\cite{Nesti:2013uwa}) and background estimations.
Thus, by taking into account those ambiguities, the least uncertain constraints are set by the diffused gamma-ray search from the dwarf Spheroidal 
galaxies by the Fermi-LAT\,\cite{Ackermann:2013yva}, which excludes the wino LSP mass below about 400GeV and in 2.2-2.5TeV.
The bino LSP is, on the other hand, free from the constraints by the indirect dark matter searches.

In Table~\ref{tab:bench mark}, we show some phenomenologically
interesting benchmark points.
Points A and C represent the bino-wino co-annihilation regions.
Since the gluino is not degenerate with the LSP, a gluino search with hard
jets and large missing energies will be effective at the LHC.  
Points B and D represent the bino-gluino co-annihilation regions.
Since the gluino is degenerate with the LSP, a search utilizing initial state radiation is necessary at the LHC~\cite{Alwall:2009zu,Bhattacherjee:2013wna}.


\begin{table}
\begin{center}
 \begin{tabular}{|c|c|c|c|c||c|c|c|}
\hline
&$m_{3/2}$ (TeV) & $L/m_{3/2}$ & $N_{\rm eff}$ & $\gamma$ & $M_3$ (GeV) &
  $M_2$ (GeV) & $M_1$ (GeV)\\
\hline
A&$50$           & $0$         & $4.5$         & $0.85$ &
$1515$ & $746$  & $729$ \\  
B&$50$           & $1$         & $5$         & $0$ &
$955$ & $993$  & $877$ \\  
C&$100$           & $0$         & $4.5$         & $0.85$ &
$2878$ & $1482$ & $1470$ \\  
D&$100$           & $1$         & $5$         & $0$ &
$1808$ & $1972$  & $1768$ \\  
\hline
 \end{tabular}
\caption{Some phenomenologically interesting benchmark points.}
\label{tab:bench mark}
\end{center}
\end{table}

\section{Discussion and conclusion}
In this paper, we have investigated the gaugino masses in
pure gravity mediation models.
We have especially focused on the threshold correction from an additional
vector-like matter, which gives a rich
structure to the gaugino masses.

It is possible that the gluino is much lighter than in the anomaly mediated gaugino spectrum, which affects
the gluino searches at the LHC.
Lighter gluino masses enhances the detectability of the gluino at the
LHC experiments.
If the gluino is degenerate with the LSP, a search utilizing initial
state radiation is necessary~\cite{Alwall:2009zu,Bhattacherjee:2013wna}.

The rich structure also allows for new dark matter possibilities such as bino-wino co-annihilation, bino-gluino
co-annihilation, or even wino-gluino co-annihilation.
It is possible that the bino is dark matter and is difficult to be
observed through direct/indirect detections, because the bino
interactions are suppressed.
If the bino decays through an $R$ parity violating interaction, however, it can provide signals
in cosmic rays.
For example, a decaying bino yields positron signals consistent with
the anomalous results~\cite{Ibe:2013jya}
reported by the PAMELA~\cite{Adriani:2008zr}, Fermi-LAT~\cite{FermiLAT:2011ab} and AMS-02~\cite{Aguilar:2013qda} 
collaborations.%
\footnote{
In Ref.~\cite{Ibe:2013jya}, the argument is made based on the wino dark
matter scenario.
As for a decay through an $R$ parity violation by $LLE^c$ operator, the
phenomenology is essentially the same as in the bino dark matter case.
}

We have also noted that the corrections to the gaugino masses in general have different phases
than the contributions from AMSB i.e.
when the corrections are from
vector-like matter as heavy as the gravitino.
We have pointed out that the phases are important phenomenologically.
For example, in the bino-wino co-annihilation region, the gluino masses
are strongly dependent on the phases, as can be seen from Figures~\ref{fig:50TeV}
and \ref{fig:100TeV}.

If the mentioned above rich structures are confirmed,
especially for a gluino much lighter than in the anomaly mediated gaugino spectrum, 
it suggests
the existences of additional vector-like matter.
It is remarkable that one can probe higher energy physics by measuring the gaugino
masses.
In particular, when the vector-like matter obtains a Dirac mass from $R$ symmetry breaking
and hence are as heavy as the gravitino, they may be found in future collider
experiments of ${\cal O}(100)$ TeV.

\section*{Acknowledgments}
The authors thank Shigeki Matsumoto for useful discussion, and Jason
L. Evens for careful reading of the manuscript.
This work is supported by Grant-in-Aid for Scientific research from the
Ministry of Education, Science, Sports, and Culture (MEXT), Japan, No.\ 22244021 (T.T.Y.),
No.\ 24740151 (M.I),  and also by World Premier International Research Center Initiative (WPI Initiative), MEXT, Japan.
 The work of K.H. is supported in part by a JSPS Research Fellowships for Young Scientists.

\appendix

\section{Review on gaugino masses from axion model}
\label{sec:axion}
In this section, we review the contribution of axion models to
the gaugino masses, following Ref.~\cite{Nakayama:2013uta}.
In general supersymmetric axion models, there are
axion chiral multiplets, which couple to
vector-like matter fields.
Since the axion multiplets are flat directions and hence not fixed, they generally obtain non-zero $F$
terms.
Thus, the gaugino masses receive threshold corrections from the
vector-like matters.
 
\subsection{KSVZ type models}
Let us consider the so-called KSVZ~\cite{Kim:1979if,Shifman:1979if} type
 axion model in which
the anomaly of the Peccei-Quinn (PQ) symmetry~\cite{Peccei:1977hh,Weinberg:1977ma,Wilczek:1977pj} of the QCD
is
mediated by additional SSM charged matters $Q$ and $\bar{Q}$.
Here, we assume a super-potential,
\begin{eqnarray}
W = \lambda X(\psi \bar{\psi}-v^2) + y \frac{\psi^n}{\mpl^{n-1}}Q\bar{Q},
\end{eqnarray}
where $X$, $\psi$, $\bar{\psi}$ are fields which have (PQ,$R$) charges $(0,2)$,
$(1,r_\phi)$ and $(-1,-r_\phi)$, respectively.
Without loss of generality, we
take $\lambda$, $y$ and $v$ to be positive and real by field redefinitions.
We assume that the axion multiplet is the only flat direction,
$\lambda v \gg m_{3/2}$.
We also assume that $y \vev{\psi}^n/\mpl^{n-1}\gg m_{3/2}$.

The scalar potential of the scalar components of $X$, $\psi$, and
$\bar{\psi}$ is given by
\begin{eqnarray}
 V &=& \lambda^2|\psi\bar{\psi}-v^2|^2 +
  \lambda^2|X|^2\left(|\psi|^2+|\bar{\psi}|^2\right)\nonumber\\
&& + m_{3/2}^2\left(a_X |X|^2+ a_\psi|\psi|^2 + a_{\bar{\psi}}|\bar{\psi}|^2\right)
+\left(2\lambda v^2m_{3/2}X + \tilde{b} m_{3/2}^2\psi\bar{\psi} + {\rm h.c.}\right).
\end{eqnarray}
Here, we assume that $X$, $\psi$ and $\bar{\psi}$ couple to the SUSY breaking
sector only through Planck suppressed interactions,
and hence, $a_X$, $a_{\psi}$ and $a_{\bar{\psi}}$ are at largest ${\cal O}(1)$. 
It should be noted that the $\tilde{b}$ term, $\tilde{b} m_{3/2}^2
\psi\bar{\psi}$ with $\tilde{b} ={\cal O}(1)$, can arise from the $R$ symmetry breaking
effect~\cite{Inoue:1991rk,Casas:1992mk} because the combination $\psi\bar{\psi}$ is
neutral under the PQ and $R$ symmetry.
As we will see, however, the $\tilde{b}$ term does not
affect gaugino masses.

The minimum of the potential is given at
\begin{eqnarray}
\label{eq:vev axion}
 \vev{X} &=& -\frac{2 m_{3/2}v^2}{\lambda\left(|\vev{\psi}|^2 +
					|\vev{\bar{\psi}}|^2\right)}\left(1 +
 {\cal O}\left(\frac{m_{3/2}^2}{\lambda^2v^2}\right)\right),\nonumber\\
 \vev{\psi} &=& \left(
\frac{a_{\bar{\psi}}m_{3/2}^2 + \lambda^2 |\vev{X}|^2}{a_{\psi}m_{3/2}^2 + \lambda^2 |\vev{X}|^2}
\right)^{1/4}v \left(1 +
 {\cal O}\left(\frac{m_{3/2}^2}{\lambda^2v^2}\right)\right),\nonumber\\
 \vev{\bar{\psi}} &=& \left(
\frac{a_{\psi}m_{3/2}^2 + \lambda^2 |\vev{X}|^2}{a_{\bar{\psi}}m_{3/2}^2 + \lambda^2 |\vev{X}|^2}
\right)^{1/4}v\left(1 +
 {\cal O}\left(\frac{m_{3/2}^2}{\lambda^2v^2}\right)\right).
\end{eqnarray}
Here, we take $\vev{\psi}$ to be positive and real by field
redefinitions.
Note that at the leading order in $m_{3/2}/(\lambda v)$, the vacuum expectation values do
not depend on the $\tilde{b}$ term. This is because the direction
$\psi\bar{\psi}$ is fixed by the super-potential.

In order to calculate the gaugino masses, let us calculate the $b$ term of
$Q\bar{Q}$.
It is given by
\begin{eqnarray}
\label{eq:b term axion}
  {\cal L}_{b-{\rm term}} &=& y \frac{\vev{\psi}^n}{\mpl^{n-1}} m_{3/2}A \bar{A}
   + n y\frac{\vev{\psi}^{n-1}}{\mpl^{n-1}} \vev{F_\psi}
				    A \bar{A} + h.c.,\\
\label{eq:F term axion}
 F_\psi &=& -\left(W_{\psi^\dag}^\dag + m_{3/2} \psi\right) = - \lambda
  X^\dag \bar{\psi}^\dag - m_{3/2} \psi,
\end{eqnarray}
where $A$ and $\bar{A}$ are the scalar components of $Q$ and $\bar{Q}$,
respectively, as in the main text.

When we calculate the gaugino masses via the $Q\bar{Q}$ loop, the
contributions from the first term in Eq.~(\ref{eq:b term axion}) cancel
with the AMSB contribution.%
\footnote{This cancellation happens only when $y
\vev{\psi}^n/{\mpl^{n-1}}\gg m_{3/2}$. For gaugino masses with
$y\vev{\psi}^n/{\mpl^{n-1}}\sim m_{3/2}$, see Sec.~\ref{sec:vector}.}
This is the famous decoupling of heavy vector-like matter~\cite{Giudice:1998xp}.
The contributions from the second term, on the other hand, do not
cancel, which lead to
corrections to the gaugino masses given by
\begin{eqnarray}
\label{eq:gaugino mass axion pre}
 \Delta m_\lambda &=& -\frac{ny \vev{\psi}^{n-1}
  \vev{F_\psi}/\mpl^{n-1}}{y \vev{\psi}^n m_{3/2}/\mpl^{n-1}}\times
  \frac{g^2}{16\pi^2} 2 C_Q m_{3/2} = \frac{g^2}{16\pi^2}2 C_Q
  \times\frac{-n \vev{F_\psi}}{\vev{\psi}},
\end{eqnarray}
where $C_Q$ is a Dynkin index of $Q$, which is normalized to be $1/2$
for a fundamental representation.

From Eqs.~(\ref{eq:vev axion}) and (\ref{eq:F term axion}), the $F$ term
of $\psi$ is given by
\begin{eqnarray}
\label{eq:F term axion result}
 F_{\psi} = - m_{3/2}\vev{\psi} \frac{a_{\bar{\psi}}-a_{\psi}}{a_{\bar{\psi}}+a_{\psi} + 2
\lambda^2 |\vev{X}|^2/m_{3/2}^2} \left(1 +
 {\cal O}\left(\frac{m_{3/2}^2}{\lambda^2v^2}\right)\right)\equiv - m_{3/2} \vev{\psi} \epsilon,
\end{eqnarray}
where $\epsilon$ is order one, unless the soft squared mass terms of
$\psi$ and $\bar{\psi}$ accidentally coincide with each others.

By substituting Eq.~(\ref{eq:F term axion result}) into Eq.~(\ref{eq:gaugino mass axion pre}), we obtain the contribution from the axion model to gaugino
masses,
\begin{eqnarray}
\label{eq:gaugino mass axion}
 \Delta m_\lambda = \frac{g^2}{16\pi^2}2C_Q n\epsilon m_{3/2}
\end{eqnarray}
Note that the phase is aligned with the AMSB contribution. 
This is because the phase of $\vev{\psi}$ and $\vev{F_\psi}$ are aligned
with each others.

Let us comment on the case with several flavors of vector-like matters,
as is the case with axion models presented in
Ref.~\cite{Harigaya:2013vja}.
Even if there are several flavors of vector-like matters, we can always
diagonalize their mass matrix.
Each mass eigenstates contribution to the gaugino masses is as given in Eq.~(\ref{eq:gaugino mass axion}).
The gaugino masses are simply multiplied by the number of the flavors.

\subsection{The SSM gaugino masses}

\begin{figure}[tb]
\begin{center}
  \includegraphics[width=0.62\linewidth]{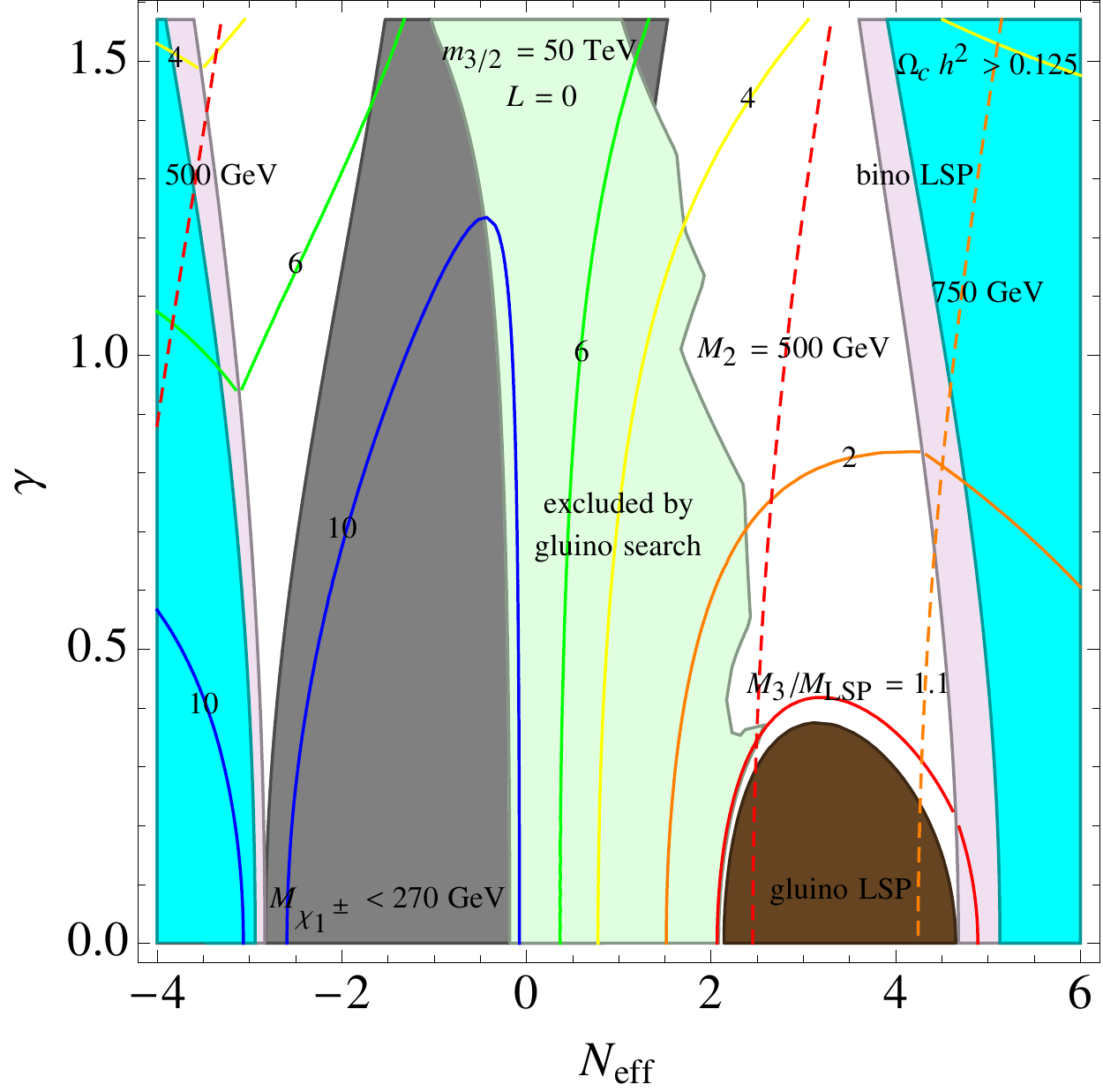}
  \includegraphics[width=0.62\linewidth]{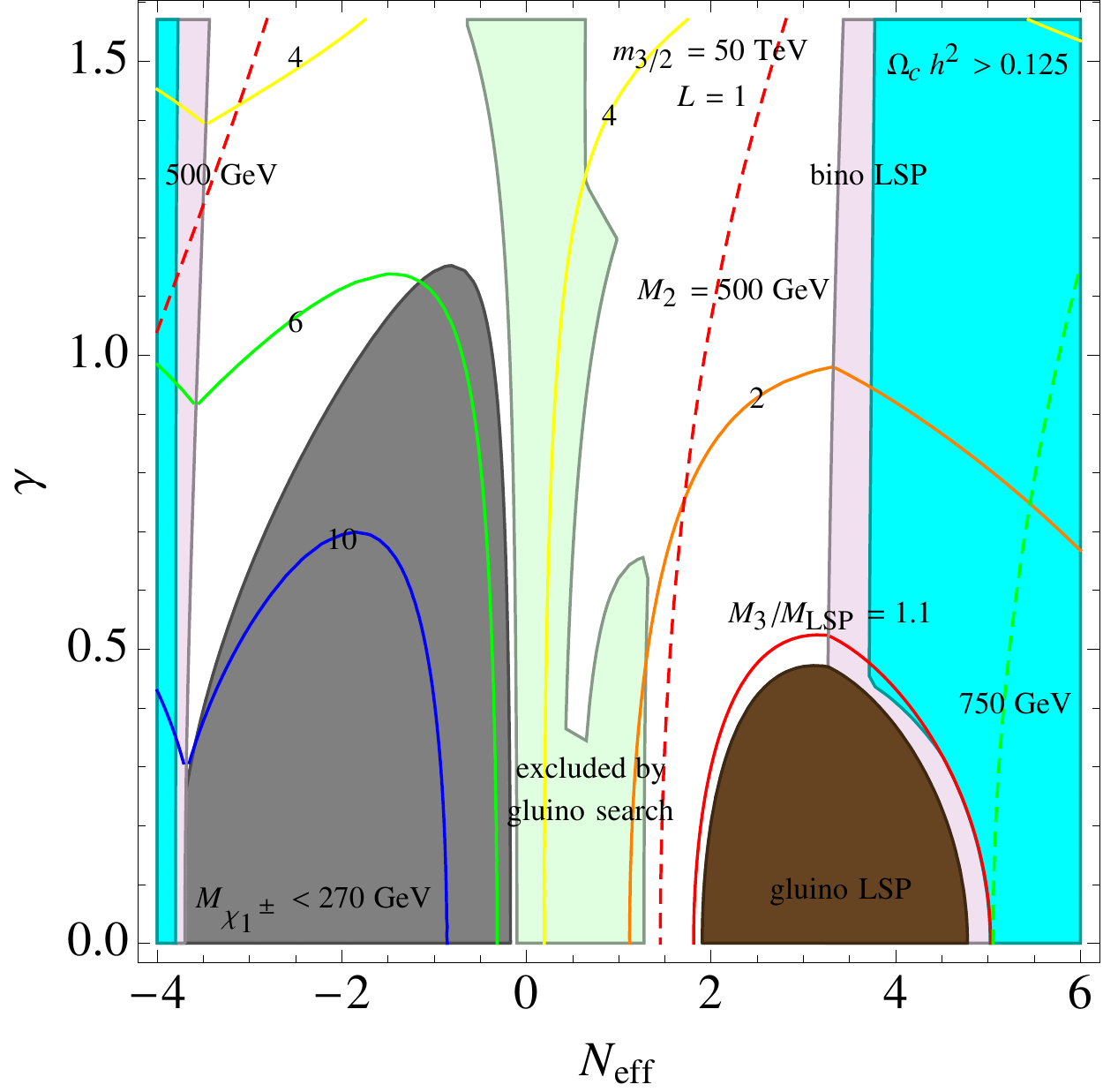}
\end{center}
\caption{\sl \small
The gaugino masses as functions of $N_{\rm eff}$ and $\gamma$
for a given $m_{3/2}$ and $L$.
The solid lines show the ratio between $M_{3}/M_{\rm LSP}$.
The dashed lines show the wino mass.
The LSP abundance is larger than the observed abundance
in the light shaded regions (light-blue), where the LSP is the bino.
In the pink shaded regions, the LSP is the bino but 
the relic abundance does not exceed the observation due to coannihilation
effects.
Here, we have neglected the effect of the Sommerfeld enhancement for simplicity
(see footnote 8).
The gluino is the LSP in the brown region.
The wino LSP has been excluded in the dark shaded regions by
disappearing track searches\,\cite{charged-wino}.
The green shaded regions are excluded by gluino searches\,\cite{gluino}.
}
\label{fig:50TeV}
\end{figure}

\begin{figure}[tb]
\begin{center}
  \includegraphics[width=0.62\linewidth]{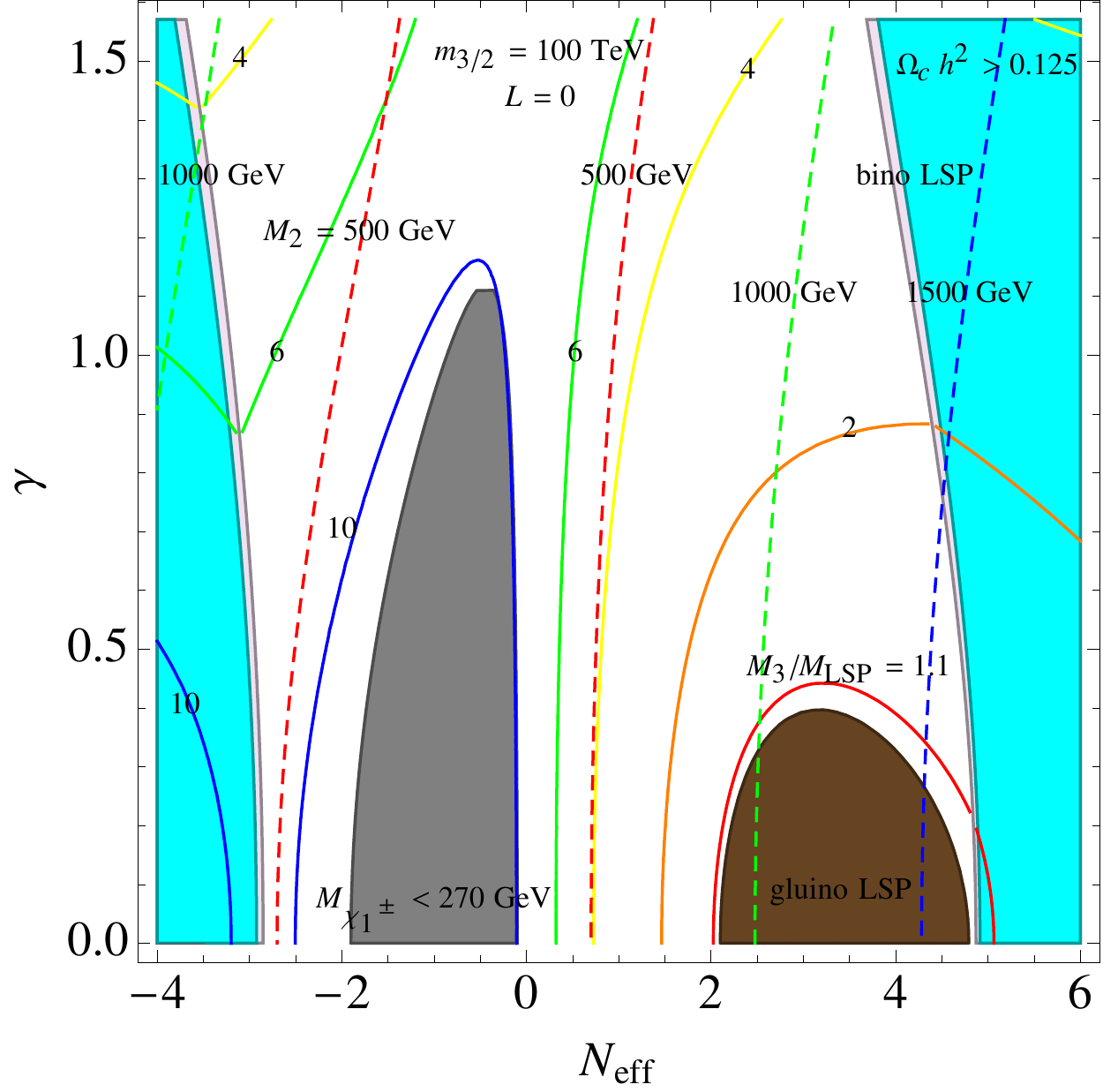}
  \includegraphics[width=0.62\linewidth]{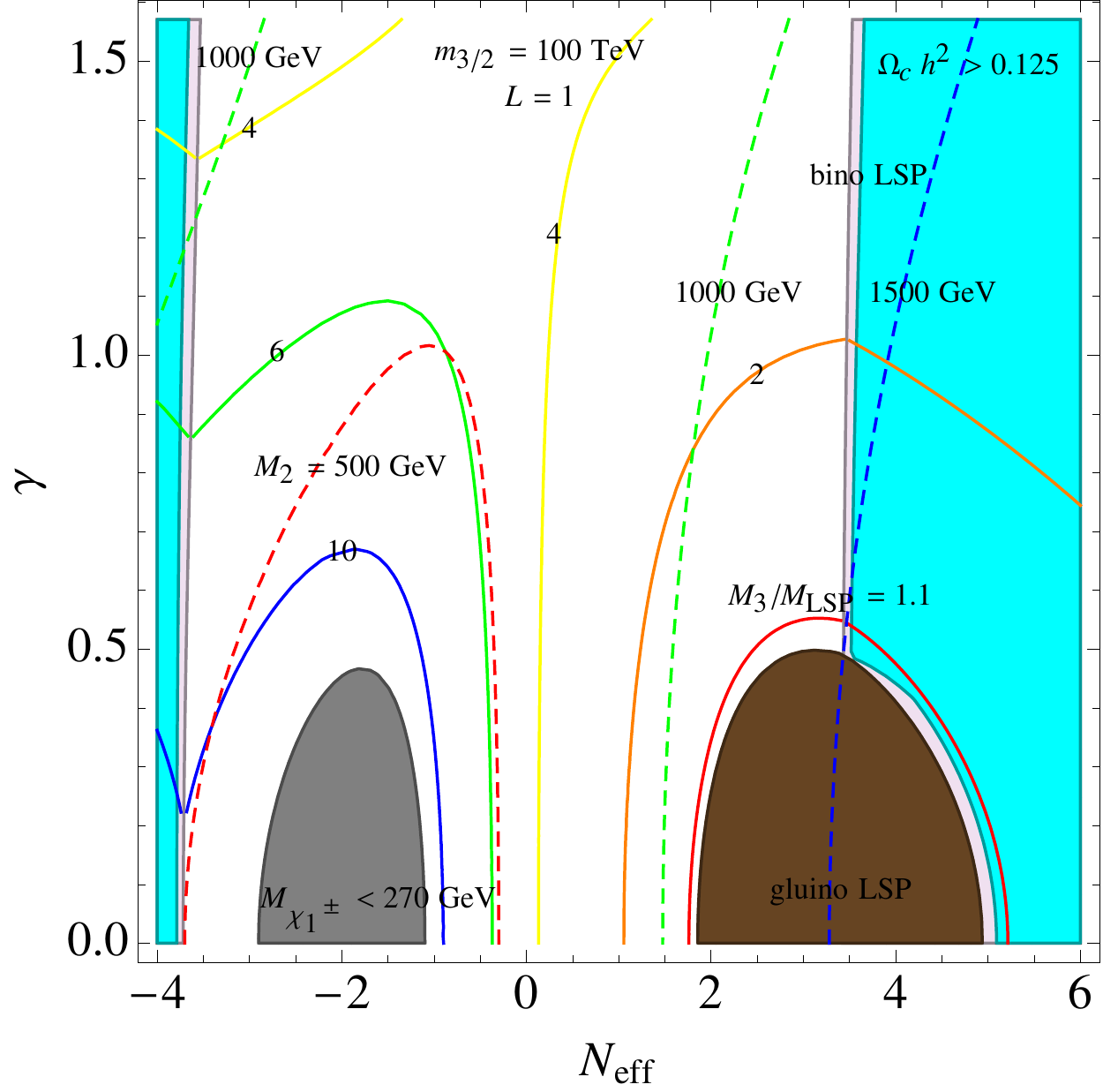}
\end{center}
\caption{\sl \small
The same as Figure\,\ref{fig:50TeV} but
for $m_{3/2}=100$ TeV and $L=0$, $m_{3/2}$.
Here, we have neglected the effect of the Sommerfeld enhancement for simplicity.
}
\label{fig:100TeV}
\end{figure}

\begin{figure}[tb]
\begin{center}
  \includegraphics[width=0.62\linewidth]{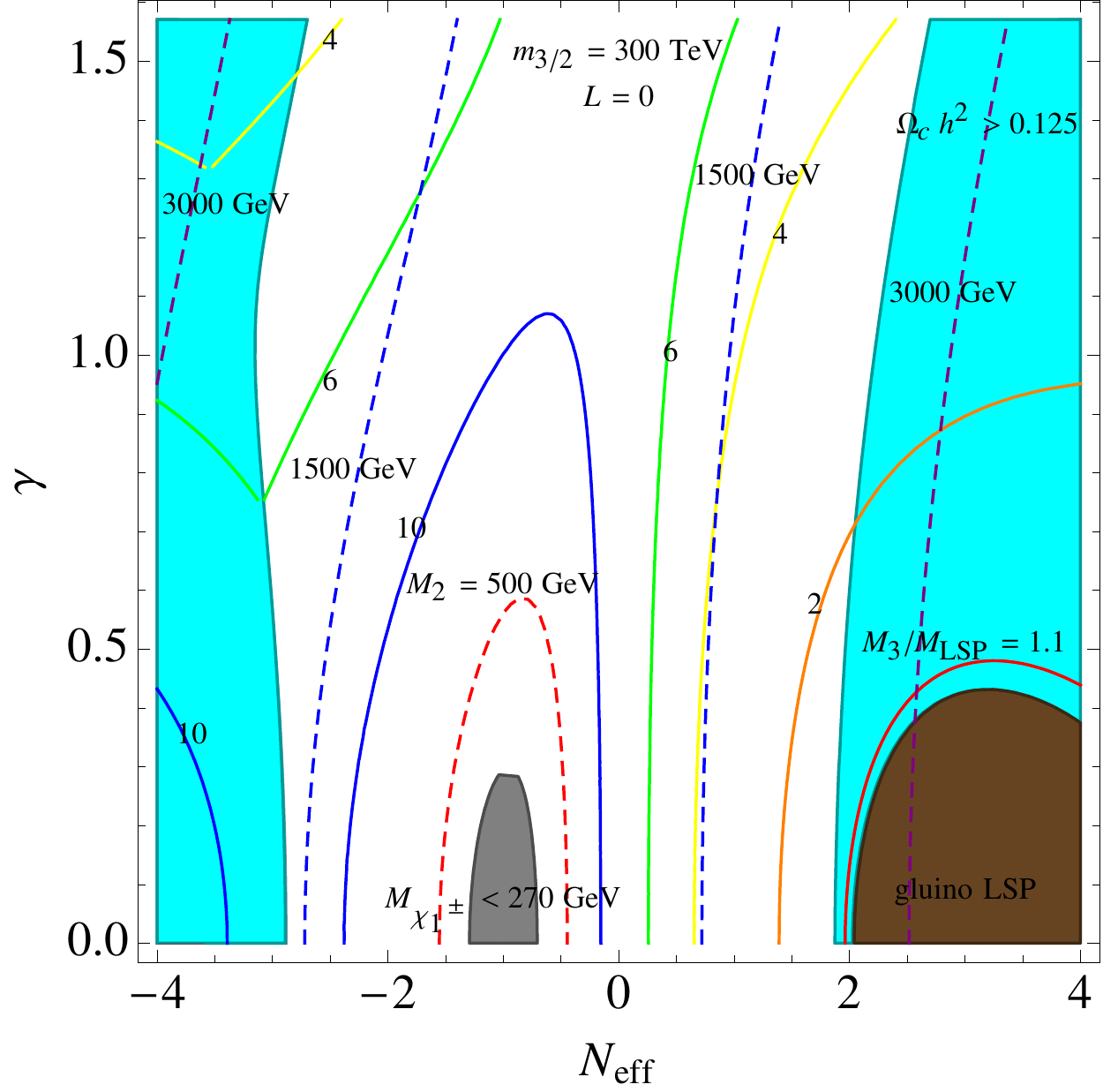}
\end{center}
\begin{tabular}{cc}
\begin{minipage}{0.5\hsize}
\begin{center}
  \includegraphics[width=1.\linewidth]{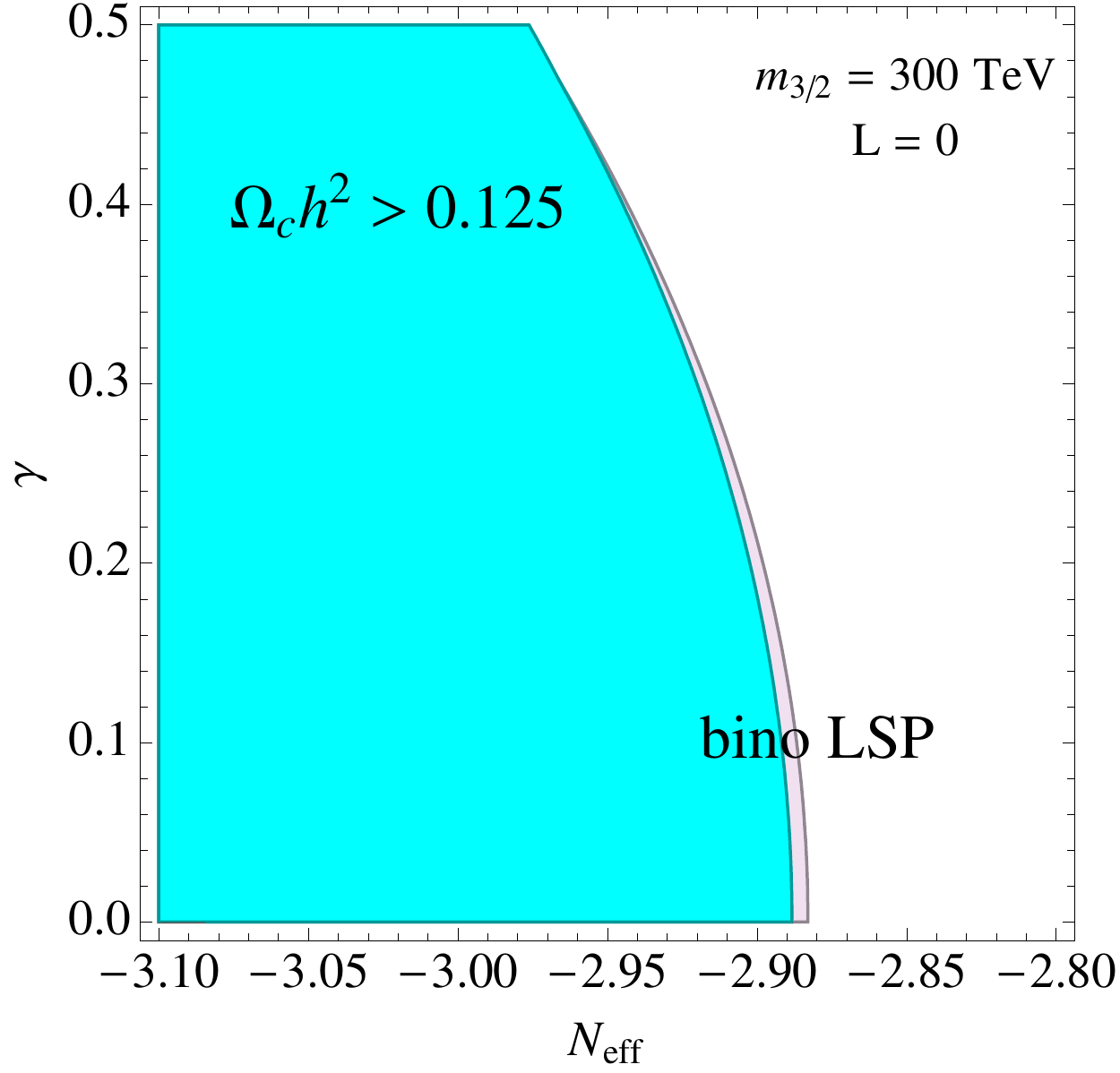}
\end{center}
\end{minipage}
 &
\begin{minipage}{0.5\hsize}
\begin{center}
  \includegraphics[width=1.\linewidth]{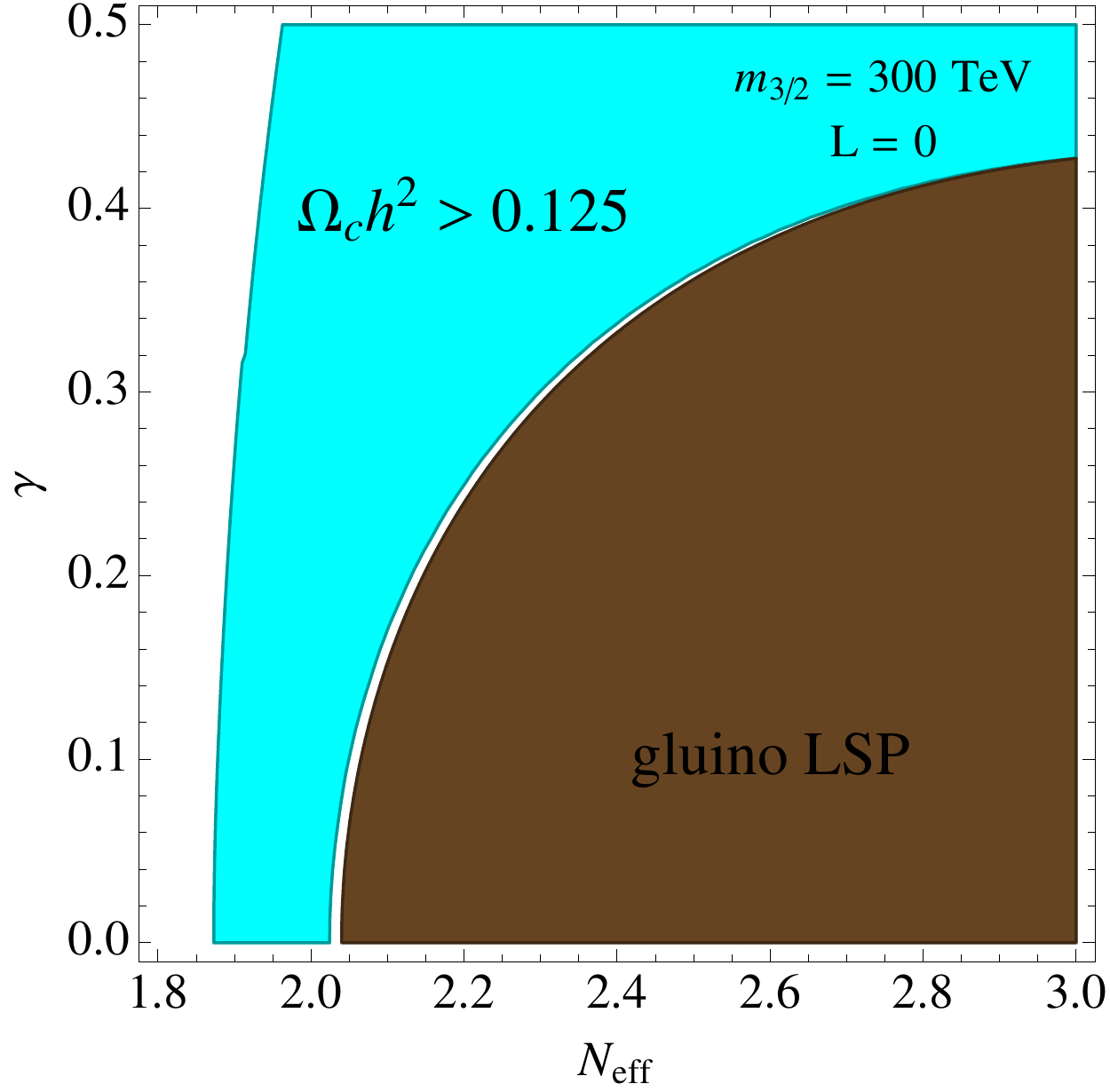}
\end{center}
\end{minipage}
\end{tabular}
\caption{\sl \small
The same as Figure\,\ref{fig:50TeV} but
for $m_{3/2}=300$ TeV and $L=0$.
In the light shaded region (light-blue) in the right hand side,
the LSP is mostly wino unlike the previous two figures.
The closeup view of the gluino-wino coannihilation region
is shown in the lower right panel.
The lower left panel is the closeup view of the wino-bino coannihilation region.
Here, we have neglected the effects of the Sommerfeld enhancement for simplicity.
}
\label{fig:300TeV}
\end{figure}

In the presence of the axion model described above, the gaugino masses
receive threshold corrections at the scale of the mass of $Q\bar{Q}$.
However, $m_\lambda/g^2$ is a renormalization invariant in
supersymmetric theory at an one-loop level.
Hence, it is not necessary to solve the renormalization equations from the
mass scale of $Q\bar{Q}$ to the gravitino mass scale for an one-loop analysis.
We can treat the corrections given by Eq.~(\ref{eq:gaugino mass axion}) as if it is
generated at the gravitino mass scale, and solve the renormalization
equations (\ref{eq:renormalization eq}).
Therefore, in this axion model, gaugino masses are given by the upper
left panel of Fig.~\ref{fig:vector-mass}.
Phenomenology can be read off from the line $\gamma=0$ of Figures
\ref{fig:50TeV}, \ref{fig:100TeV} and \ref{fig:300TeV}. 
In axion models with a large number of additional
matter~\cite{Harigaya:2013vja}, $N_{\rm eff}$
would be considerable.

\newpage
\thispagestyle{headings}

\end{document}